# Quad-Core RSA Processor with Countermeasure Against Power Analysis Attacks

Javad Bagherzadeh, Vishishtha Bothra, Disha Gujar, Sugandha Gupta, Jinal Shah
Department of Electrical Engineering and Computer Science (EECS)
University of Michigan, Ann Arbor

*Abstract*— Rivest-Shamir-Adleman (RSA) cryptosystem uses modular multiplication for encryption and decryption. So, performance of RSA can be drastically improved by optimizing modular multiplication. This paper proposes a new parallel, high-radix Montgomery multiplier for 1024 bits multi-core RSA processor. Each computation step operates in radix 4. The computation speed is increased by more than 4 times. We also implement a True Random Number Generator based resilience block to protect the coprocessor against power attacks.

## I. INTRODUCTION

In the mid 1970's early public key cryptosystems were developed to provide a solution for two parties to communicate in a secure manner over an insecure channel. Public key cryptosystems are used to establish secure communication and to provide non-repudiation through the use of digital signatures. RSA cryptosystem, named after its creators, is one of the most popular public key cryptosystems. The RSA cryptosystem has been utilized for e-commerce, various forms of authentication, and virtual private networks. The importance of high security and faster implementations paved the way for RSA crypto-accelerators, hardware implementations of the RSA algorithm.

RSA uses Montgomery multiplication for modular exponentiation. The main complexity of modular multiplication lies in a series of lengthy operations. Several Montgomery multiplication methods have been proposed and analyzed for area and speed constraints [1], but many of them cannot be implemented in hardware. The method for parallel k-partition Montgomery Multiplier has been proposed which can be implemented in hardware [3]. Using high radix digit set is a well-known technique to implement arithmetic algorithms in hardware. Dividing the multiplier into partitions, reduces the number of partial products and hence reduces the number of clock cycles required to complete the task. However, the additional complexity introduced for selecting partitions of the multiplier limits the computation time. Hence, a trade-off between area, speed, cost and energy consumptions needs to be achieved.

Many public key cryptosystems base their security on the difficulty of solving a known hard problem such as discrete logarithms or integer factoring. Thus, they are assumed to be very secure; however, many researchers have shown that breaking cryptosystems is not necessarily as difficult as solving these known hard problems. Recently, several side channel attacks have been investigated which substantially reduce the mean time to disclosure of the secret key [2][11]. Although computation power has increased with Moore's law[23][24], the large increase in computation costs associated public key cryptosystems has put a significant strain on available computing resources. Thus, there is a growing need for hardware acceleration of public key cryptosystems to reduce the burden of using them. Public key cryptosystems have become ubiquitous in computing devices. Not only have servers but also embedded systems have forced designers to add additional hardware for cryptosystems referred as crypto-accelerators. Crypto-accelerators are very promising as they typically achieve better performance and better power efficiency than a software implementation on a generic processor.

In this paper, we intend to implement and analyze a power attack resilient quad-core ASIC RSA processor with parallel high-radix Montgomery multiplication in hardware. The power and area consumption of the four core architecture will be evaluated. Power-attack resilience will be measured by correlating data and power.

*A. RSA algorithm*

The RSA cryptosystem is based on some mathematical functions which are easy to compute in one direction and difficult to compute in the opposite direction without special information. These kind of functions are called trapdoor one-way functions. In case of RSA, the idea is that multiplication is relatively easier than factorization. Multiplication can be computed in polynomial time whereas factoring time can grow exponentially proportional to the size of the number.

RSA algorithm consists of typically 3 steps - key generation, encryption and decryption [4]. Public key (n, e) is published for everyone and private key (p, q, d) must be kept secret. Then by using these keys encryption, decryption, digital signing and signature verification are performed. RSA algorithm modular length is commonly 1024 bit or even 2048 bit to ensure safety. Therefore, generation of public and private key, encryption, decryption all rely on the high-speed of computer.

The RSA algorithm requires the computation of modular exponentiation, which is broken into a series of modular multiplications [1] as shown in Algorithm 1: In 1985, Montgomery introduced a new method for modular multiplication. The Montgomery multiplication algorithm is used to speed up the modular multiplications and squaring required during the exponentiation process in RSA. The RSA algorithm here is the commonly used R-L Algorithm as shown in Algorithm 2. P is plaintext, E is the exponent, M is the modulus, C is the constant $2^{2n}$ (mod M) (which must be precomputed), Mont () is Montgomery multiplication and R is the result.

## B. MONTGOMERY MODULAR MULTIPLICATION

The Montgomery algorithm [2][5] of two numbers, say, a and b computes the following:

$$r = Mont(a,b) = a.b.r^{-1} \mod M$$

Where $r = 2^{*k}$ (k = no. of bits of M). The a and b M-residue multiplicands are obtained from real numbers a and b using the following:

$$A = Mont(a, r^2) = a.r \mod M$$

$$B = Mont(b, r^2) = b.r \mod M$$

Thus, the Montgomery multiplication of two M-residue numbers are computed as,

$$R = Mont(A,B)$$

$$R = (a.r).(b.r).r^{-1} \mod M = a.b.r \mod M$$

The result is also an M-residue number and needs to be converted back to a real number in the following way,

$$r = Mont(R, 1) = a.b.r.r^{-1} \mod M = a.b \mod M$$

The Montgomery approach is very efficient as it avoids the time consuming trial division which is a bottleneck for most other algorithms. Thus, it forms the basis of many implementations of modular multiplication, both in software and hardware. The Montgomery algorithm computes the result by replacing the division operation of k times division by a power of 2 division. Thus, not only the computation time but also the area is reduced in hardware implementations. Radix-2 Montgomery Modular Multiplication can be seen in Algorithm 3 [6].

---

**Algorithm 1**: RSA using Modular Exponentiation

// Compute $y = x^d \pmod N$
// where, in binary, $d = (d_0, d_1, d_2, ..., d_n)$ with $d_0 = 1$
$s = x$
**for** $i = 1$ to $n$
    $s = s^2 \pmod N$
    **if** $d_i == 1$ **then**
        $s = s.x \pmod N$
    **end if**
**next** $i$
    return $s$

---

**Algorithm 2**: RSA using Montgomery Exponentiation

RSA (P, E, M)
{
    $e = 2^{2n} \pmod N$;
    P = Mont (e, P, M);      (Mapping)
    R = Mont (e, 1, M);
    **for** i = 0 to k – 1 **do**
        **if** ($e_i == 1$) **then**
            R = mont (R, P, M); (Multiply)
        **end if**
        P = mont (P, P, M);    (Square)
    **end for**
    R = mont (1, R, M);      (Remapping)
    Return R;
}

---

**Algorithm 3**: Radix - 2 Montgomery Multiplication

**Require** odd M, $n = 1 + [\log_2 M]$,
    $X = \sum_{i=0}^{n-1} x_i 2^i$, $Y = \sum_{i=0}^{n-1} y_i 2^i$,
    with $0 \leq X, Y < M$
**Ensure** $Z \equiv XYR^{-1} \pmod M$, with $0 \leq Z < M$

1.    $S[0] \leftarrow 0$
2.    **for** $i \leftarrow 0$ to $n - 1$ step 1 **do**
3.        $a \leftarrow S[i] + x_i Y$
4.        $S[i + 1] \leftarrow (a + a_0 M)/2$
5.    **end for**
6.    **if** $S[n] \geq M$ **then**
7.        $S[n] \leftarrow S[n] - M$
8.    **end if**
9.    **return** $Z \leftarrow S[n]$

---

Radix 2 means at each loop iteration in lines 2-5, one bit of X is multiplied with Y.

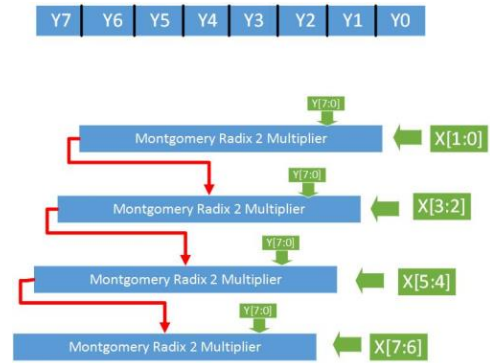

Fig. 1. Radix-4 Montgomery Multiplier (MM)

This operation is similar to multiplying with serial multiplication. In application with higher speed requirement,

higher radix MM is used in which multiple bits of X is multiplied with Y which can be seen Fig. 1. Consequently, the number of running iterations in the loop would decrease by using a higher radix MM which needs more area. Different MM with varied radixes will have different speed and area, so appropriate specification needs to be determined based on the design constraints.

A way to speed up the Montgomery Multiplication by distributing the multiplier operand bits into k-partitions is proposed in [6]. All of them process in parallel and use an identical algorithm. Each partition executes its task in = n/k steps. Even though the computation step operates in radix, the complexity is reduced by the use of a limited digit set [6].

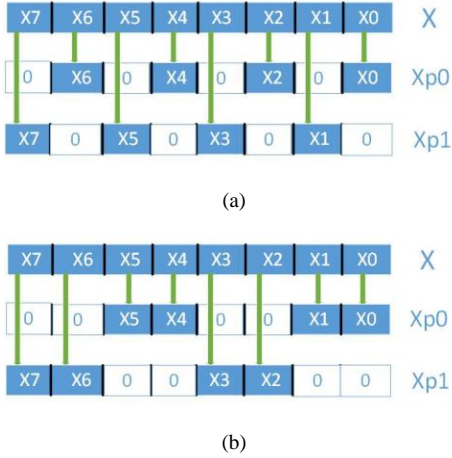

(a)

(b)

Fig. 2. The distribution of bits of into two decomposed multiplier operands
(a) Radix-2 (b) Radix-4

---

**Algorithm 4**: Montgomery Multiplication Partition j (MMP)

***Require*** odd $M$, $n = 1 + [log_2 M]$,

$X = \sum_{i=0}^{n-1} x_i 2^i$, $Y = \sum_{i=0}^{n-1} y_i 2^i$,

with $0 \leq X, Y < M$, $j^{th}$ - partition

$0 < k, t < n$, $kt = n$, $k$ partitions

***Ensure*** $Z_{Pj} \equiv MMP (j, X, Y, M) \equiv$

$(X_{Pj} Y R^{-1})$ mod $M$, with $0 \leq Z_{Pj} < M$,

where $X_{Pj} = \sum_{i=0}^{n/k-1} x_{j+ik} 2^{j+ik}$

1.     $S_{Pj}[0] \leftarrow 0$
2.     **for** $i \leftarrow 0$ to $n - 1$ step 1 **do**
3.         $a \leftarrow S_{Pj}[i] + x_{j+ik} 2^j Y$
4.         $q_{k-1 \ldots 0} \leftarrow a_{k-1 \ldots 0} (2^k - M^{-1}{}_{k-1 \ldots 0})$ mod $2^k$
5.         $S_{Pj}[i+1] \leftarrow (a + q_{k-1 \ldots 0} M)/2^k$
6.     **end for**
7.     **return** $Z_{Pj} \leftarrow S_{Pj}[n/k]$

---

An effective result is obtained when the manipulation of multiples of the multiplicand operand is performed by distributing the multiplier operand bits into partitions that can process them in parallel. The partitions of the original multiplier are used to express new multiplier operands that are used to perform Montgomery Multiplication in radix. Partitioning in Radix 2 and 4 can be seen in Fig. 2. The adapted MM for each partition can be seen in Algorithm 4 [6].

### C. OVERVIEW OF SIDE-CHANNEL ATTACKS

In cryptography, an attack based on the leaked information from hardware implementation is called a "side-channel attack" [2]. Active attacks, also referred as tampering attacks, require access to the internal circuitry of the attacked device [8]. In passive attacks, key dependent variation of cryptographic processor is monitored to retrieve the private key. There are mainly four types according to the type of the leaked information: Power Analysis [11], Electromagnetic Analysis [9][10], Acoustic Analysis [12] and Timing attacks[13]. All passive attacks can be either simple or differential. Power Analysis (PA) attacks are based on analyzing the power consumption of the cryptographic device while it performs encryption or decryption [11]. The physical supporting point of these attacks is that today CMOS technology is the one to be used most commonly for digital integrated circuit implementations and the dissipated power in these circuits is dependent on the transitions and activity of the circuit which is also dependent on the data that is processed.

## II. DESIGN AND IMPLEMENTATION

As discussed in the previous section, radix 2 or higher can be utilized in implementing k-partition algorithm. Basically, the k-partition method is used to divide the Montgomery multiplication to multiple partial multiplications that are performed in parallel in different cores. But the radix, which is the width of operations in each core influences the architecture, area and speed of each core and consequently this has to be designed carefully. Primarily, radix is the number of bits of X that are multiplied by Y in each iteration and determines the size of intermediate values and multipliers.

On the other hand, as Y is constant in all iterations and all cores, it is possible to pre-compute the results of multiplications in the beginning of the process and just use the final results to reduce the critical path and computation time. For example, by choosing radix 8, it is necessary to compute the product of Y with 011, 101, 111 and all others can be generated by shifting Y or the mentioned products. So the first decision for choosing the radix value was $2^4$ which seemed reasonable in terms of speed and area. But after the implementation and APR, the area for this design turned out to be much larger than expected. So the radix was reduced to $2^2$. By this change, the number of iterations to compute the Montgomery product will increase from 64 in radix $2^4$ to 128 iteration in radix $2^2$. In the same frequency, the higher radix would be almost two time faster than the low radix one. However, the width of multipliers and adders in the data path of $2^2$ design would be half of the $2^4$ one and consequently the critical path is much less. It is, thus, possible to increase the frequency for 128 cycles of computation. So the penalty in throughput would not be much although we are considerably reducing the area.

## A. TOP level: RSA Processor

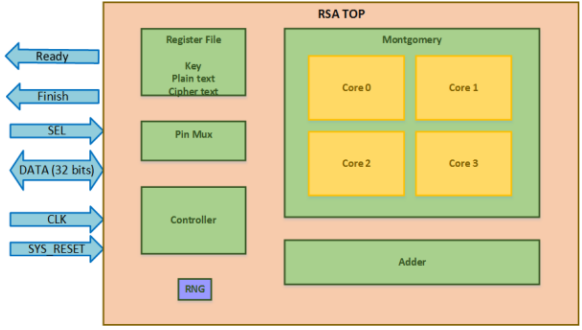

Fig 3. Top Level Architecture for RSA processor

1024-bit RSA processor is implemented using 4 identical independent cores, each computing the partial product without any inter-core data dependencies. Each 1024-bit message is divided into four partial segments of same size, each going to a core for computation. Using a higher radix (radix 4) multiplication further speeds this up. Hence, the computation speed is increased as each partition computes the partial product in 1024/8 iterations. The partial products are added appropriately to give the final modular product. The top level has four cores for partial product computations and an adder as shown in the Fig. 3.

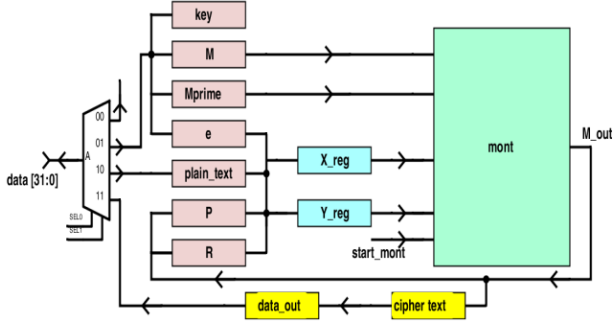

Fig 4. Top Architecture

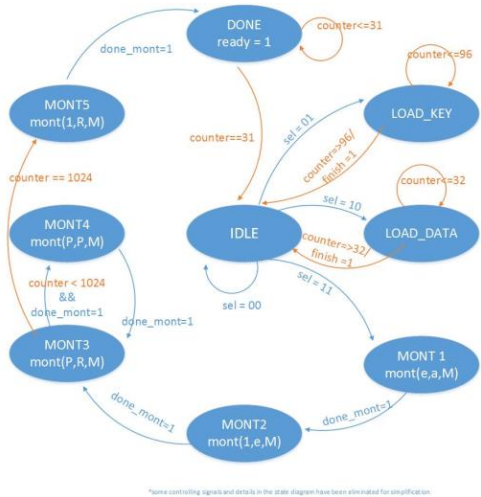

Fig 5. Finite State Machine depicting top module

**Algorithm 5.** (RSA using Montgomery Exponentiation

RSA (P, E, M)
{
  $e = 2^{2n} \pmod{M}$;
  P = Mont (e, P, M);         ← MONT 1
  R = Mont (e, 1, M);         ← MONT 2
  for i = 0 to k – 1 do
    if ($e_i$ == 1) then
      R = mont (R, P, M);     ← MONT 3
    end if
    P = mont (P, P, M);       ← MONT 4
  end for
  R = mont (1, R, M);         ← MONT 5
  Return R;
}

*1) Architecture:*
Fig. 4 shows the datapath for the top module. The values of *key, M, Mprime, e, plain_text* and *cipher_text* are pin muxed as a *data* [31:0] inout pin and are stored in their respective register through the given bus. Moreover, the required input for the Montgomery operation is provided through the registers *X_reg, Y_reg, M* and *Mprime*. The output of the Montgomery block, *M_out* is ready to sent back to *P* or *R* registers for further computation or is stored in *cipher_text* register to be sent out through the data pin.

*2) Control and timing:*
Once the *SYS_RESET* is deasserted, *sel* can be 01, 10 and 11 for loading the key, loading the data and starting Montgomery respectively.

   *a) Loading the key:*
When *sel* is 01, the state machine goes to *LOAD_KEY* state, and remains in that state for 96 cycles till the *key, M, Mprime* and *e* are loaded serially into their respective registers.

   *b) Loading the data:*
When *sel* is 10, the state machine goes to *LOAD_DATA* state for 32 cycles to store the serial 32-bit data into plaintext register. The state returns to *IDLE* after each of these states making them independent and re-executable.

   *c) Start Montgomery:*
Once the key and data are loaded, and *sel* becomes 11, the state machine goes through the various steps of RSA algorithm shown in *Algorithm 5* through the states *MONT1* to *MONT5* as depicted in the Fig. 7 once the RSA output is computed after 2051 Montgomery operations, the FSM goes to state *DONE*, resets the Montgomery block and asserts the ready output in top to indicate that the output is ready. This 1024-bit ciphertext is available as output through the pin-muxed 32-bit inout data pin serially in 32 consecutive clock cycles.

## B. Montgomery Module

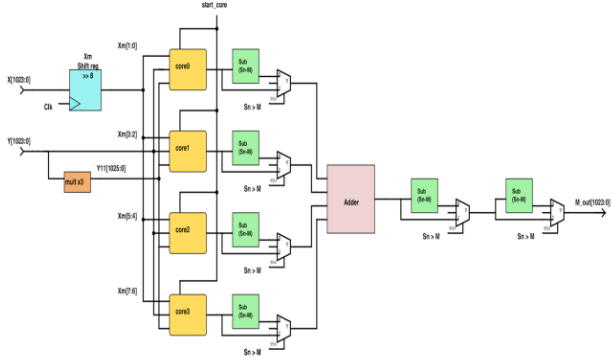

Fig 6. Mont module architecture

### 1) Architecture:

The data from the *X_reg* and *Y_reg* from top inputs into the Mont block. Multiplier X is stored in register *Xm* from where it enters the 4 cores of the Mont block by shifting 8 bits at a time which can be seen in Fig. 6. The multiplicand *Y* and *Y*3* also goes as input to the 4 cores. If the partial products of the core are greater than *M*, they are subtracted from *M*. These 4 partial products are then added together in a 1024-bit adder, and subsequently *M* is subtracted from the sum to obtain correct value for *modM*. This 1024-bit *M_out* is sent to the top as the Montgomery output.

### 2) Control and Timing:

The top asserts the *start_mont* signal to indicate the Montgomery to start. The Montgomery asserts the *start_core* signal to indicate the cores to start processing. State *INIT* moves to state *WORK*. *Xm[1:0], Xm[3:2], Xm[5:4]* and *Xm[7:6]* are connected to core 0, 1, 2 and 3 respectively. *Xm* is continuously shifted by 8 bits in each clock cycle to provide all the values to the core for computation. The Mont FSM stays in state *WORK* for 128 clock cycles which are tracked using a counter to complete the computation of the partial product. The partial product is sent out and done signal is asserted.

## C. CORE

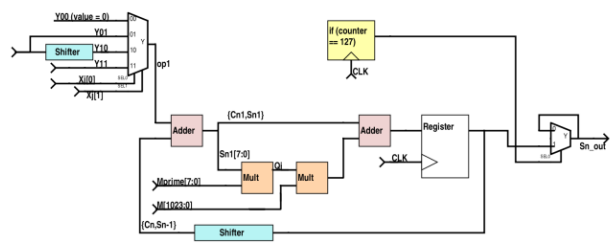

Fig 7. Core module architecture

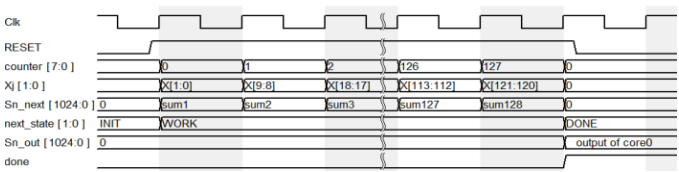

Fig 8. Timing diagram for core module

### 1) Architecture

2-bit *Xj* from the Mont block is input to the core along with *Y* and *Y*3*. To minimize the computation and time, Xj[1:0]*Y[1023:0] has been implemented using the signals *Y00, Y01, Y10, Y11* as mux input. This value is added to the value of Sn computed in the previous cycle. As shown in Fig. 7, the lower 8 bits of *Sn1* which is the output of the previous addition is multiplied to *Mprime* to generate *Qi* which is further multiplied with *M*. This output is added to *Sn1*, stored in a register and shifted by 8 bits for the next computation. After processing the entire segment in one partition for 128 cycles, the partial product of the Montgomery is passed to Montgomery module for summation.

### 2) Control and Timing

After receiving *start_core*, the core starts computation of the partial product by entering the WORK state from INIT state for 128 cycles. Once the output is computed, the FSM goes to DONE state, the done signal would be asserted and *Sn_out* is given as output of the core shown in Fig. 8.

## III. OVERCOMING POWER ATTACKS

The key of the RSA can be traced through Power Analysis and thus it is important to mask the power in the way that it does not reveal the key information.

### A. Algorithmic level

The RSA operational dependency on the key is eliminated at an algorithmic level by performing modular multiplication independent of the key value. However, the data dependent power variation still remains and is corrected using other techniques.

### B. Inherent Power resilience:

Montgomery multiplication is being carried for 2 bits in each core at each iteration. There are 4 such independent cores being executed at the same time. So, there are 8 bits being executed in a single clock cycle. So that makes it 128 times more difficult to identify the secret key compared to a single bit Montgomery with one core. Although the required complexity of power attack is increased, it is still susceptible to power attacks. Several Power resilience techniques under consideration for this project are described in the next section.

### C. Custom power resilient techniques:

#### 1) Switched Capacitor Technique:

First, Switched capacitance method as described in [7] was considered. In this technique, as shown in Fig. 9, a switched capacitor block isolates the core from the power supply and avoids any variable switching activity directly from the power supply. This is implemented using 3 arrays of capacitors which are switched between three phases:

1- Charged from the power supply
2- Provide charge to the encryption core
3- Discharged to a known value

Each capacitor is charged from the power supply and is then isolated while it provides charge to the sensitive blocks of the encryption core. The capacitor is discharged to a known value before it is recharged. This ensures that the current drawn remains constant and provide resilience. Large capacitors are needed in order to provide supply to the encryption core resulting in a huge area penalty. Also, the voltage of the capacitor providing power to the core is going to drop resulting in a performance penalty.

*2) Op-amp based Current Equalizer:*

In order to overcome the area penalty, an op-amp based current equalizer method was considered. An op-amp circuit can be used to dynamically bias an NMOS connected in path parallel to the core to make up for any mismatch in current of the core. This ensures always the same current being drawn from the power supply. Essentially core supply is maintained at constant voltage. The setup is shown in Fig. 10. Because of finite response time of op-amp to overcome the current variation by core, attackers can use this this timeframe to retrieve the secret key. So, the core is still susceptible to power attacks.

*3) Constant Current Source Operation*

Constant current based operation was also proposed. As shown in Fig. 11, first the core supply capacitance is charged to appropriate voltage and the current drawn by the core is copied and supplied back to the core to maintain constant core supply voltage. The extra current from the current source is dissipated by the resistor. This technique has very low area overhead and is able to reach very high level of power attack resiliency. But performance is severely degraded because of core supply will be at lower voltage than VDD.

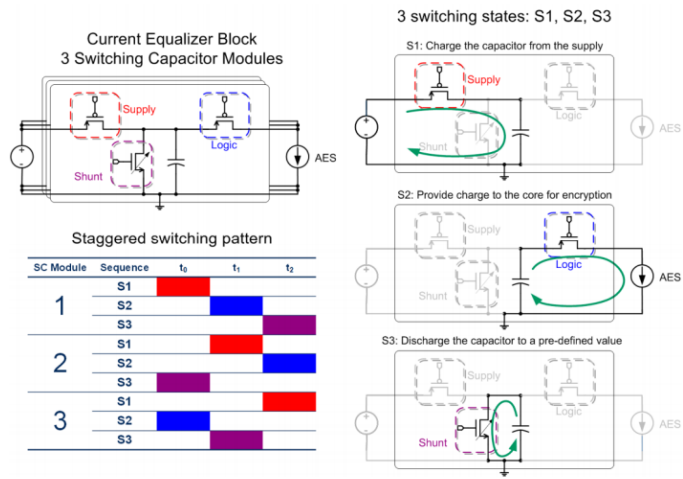

Fig 9. Switched Capacitor Technique [7]

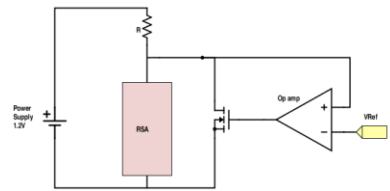

Fig 10. Current equalizer for power-attack resilience

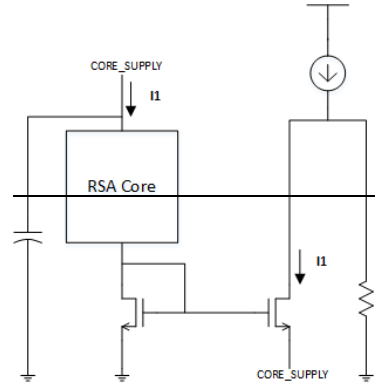

Fig 11. Constant current source circuit

*4) Proposed resilience method*

In order to overcome all limitations as mentioned above, we propose a True Random Number Based Power Resilience technique. Conventionally, Randomly generated numbers are used to turn on/off oscillators to generate random power variation. This random power variation breaks any correlation between any power model and power measurements of the coprocessor. Jitter Amplifier based random number generation is described in [14]. Jitter amplifier increases the uncertainty of the random numbers and allows us to sample it at much higher frequency. As it is true random number generator, bit stream generated after each system reset is different. Random bit stream generated by Jitter Amplifier in [15] passes all 15 tests in NIST P800-22 with 99% confidence. The complete setup is as shown in Fig. 12. Jitter amplifier uses dynamically varying delay cells to increase uncertainty and amount of Jitter variation for the system clock Now, this clock is used as a sampling clock to generate a random number bit stream . Now the delay cells in Jitter Amplifier have random power variation. In order to reduce area overhead, these delay cells itself can be used as a source of random power. As this power gets added to the core power, the resultant power waveform is also completely random for drawing random power from the power supply. This proposed technique reduced area penalty compared to conventional method as the circuit module is shared for both generating random number and create random power variations. As the critical path of the core is not affected, there is no performance penalty for this method.

IV. RESULTS AND SIMULATIONS

The MATLAB Simulations for parallel Radix-2, Radix-4 and Radix-16 were performed. The golden bricks for the core and top level are created and expected results are achieved for different radixes. Based on the first decision for radix in this project, the RTL for the radix-16 architecture of core block, Montgomery Multiplier and Top level RSA Processor was coded and debugged. The core was synthesized with no slacks or violations and post synthesize verification was also done which resulted in the correct expected output. However, after the implementation, the area for this design turned out to be much larger than expected.

So the radix was reduced to $2^2$ and once more all the steps containing RTL coding, synthesize, and APR were done and each step was verified by comparing with the respective golden brick. The synthesis reports' summary of two implemented design are mentioned in Table I.

The simulations for the switched capacitor current equalizer are performed and the correct switching between capacitor modules is obtained. The two methods mentioned for power attack resilience will be simulated against the core and considered against performance, area and power tradeoffs.

Fig. 13 shows a random bit stream generated for several clock cycles. Red waveform is random bit stream and pink waveform is the system clock. Fig. 13 shows how the Jitter of the system clock is increased after Jitter Amplifier. Fig. 14(a) shows the frequency of variation of the system clock given as input to Jitter Amplifier.

Fig. 14(b) shows the frequency variation of the clock given as output of the jitter amplifier. It can be seen that range of frequency variation is increased from 548 MHz-553 MHz to 500 MHz-625 MHz. The red waveform is the clock and the blue waveform is the frequency for each clock cycle.

TABLE I.  SPECIFICATIONS OF CORE IN RADIX $2^2$ AND $2^4$ DESIGN

| Radix | F (MHz) | Area (mm²) Synthesis | Area (mm²) APR | No. of Cells | RSA Duration (ms) | Power (uW) |
|---|---|---|---|---|---|---|
| $2^2$ | 277 | 9.9 | 2.5*2 | 8.6K | 0.945 | 360 |
| $2^4$ | 166 | 16.1 | - | 123K | 0.812 | 357 |

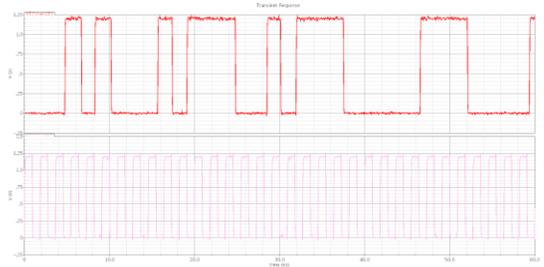

Fig 13. Random bit stream generator over different clock cycles

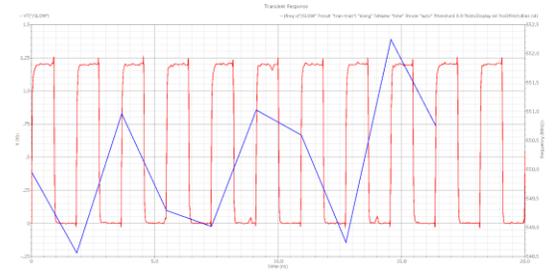

Fig. 14(a). Input Clock of Jitter Amplifier showing frequency variation

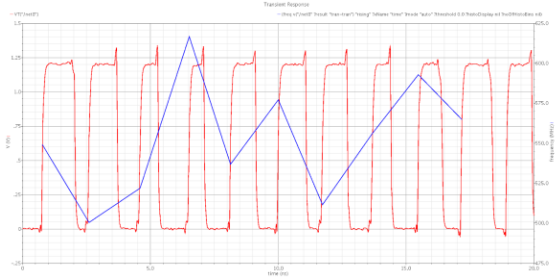

Fig. 14(b). Output Clock of Jitter Amplifier showing incresed frequency variatiion

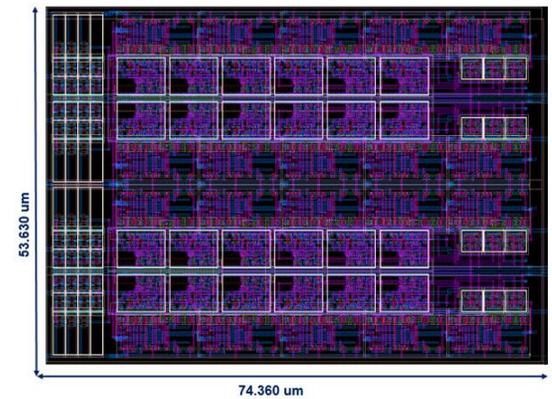

Fig. 15 Layout of resilience block

Layout of the custom block is as shown in Fig. 15. The resilience block area is 3987 um2 which is only 0.02 % of the total area. Fig 19 shows the average power of the this block for a time period one RSA instruction after 10 system resets.

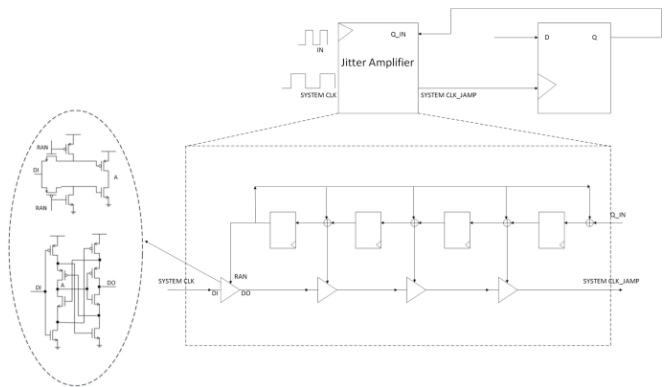

Fig 12. Proposed jitter based noise generator circuit

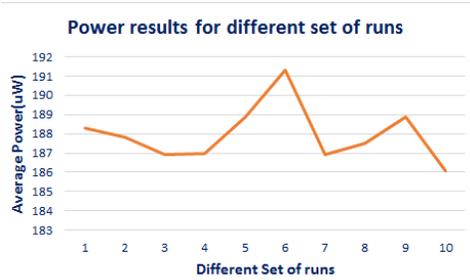

Fig. 16 Average power after different systm reset

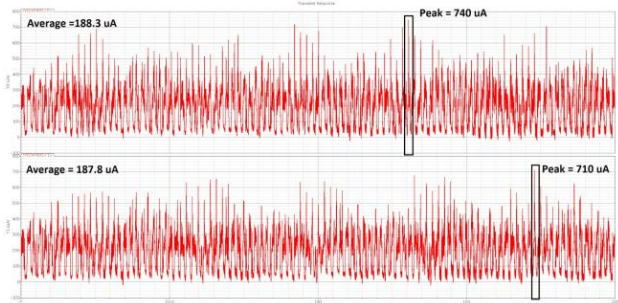

Fig. 17 Two different power traces for same duration

It can be seen that the average power is not constant showing the true randomness of the reseilience block. Fig. 16 shows two different power traces for the same time frame. It can be easily seen that the spike variation for both the run are different.

After integrating the design with the resiliency block, the efficiency of the proposed against power analysis attack needs to be verified. By referring to publication in this regard, it can be seen there is no specific metric in this purpose unless the number of required traces to measure show the unfeasibility of attack by increasing the number of traces to attack the resilient design and compare it to the non-resilient block. To obtain a single trace, we need to run post APR simulation and obtain the power trace by running the power simulation using Power Prime. As the run time for one trace of one run encryption is long and our design is very complex obtaining a single trace would take a considerable time. So using the conventional method to verify the resiliency by obtaining exhaustive number of traces is not feasible.

So here, we propose a new method to verify the design, which is based on Correlation Power Attack. CPA has been called the most efficient power analysis attack as it uses correlation coefficient to show the maximum correlation between power model of the processed data and power traces. So by implementing CPA on protected and un-protected design and comparing the result, it is possible to measure and verify the design.

For this purpose, 100 power traces of protected and un-protected design was obtained using the mentioned procedure in Power Prime and converted to usable data in Matlab. After implementing CPA in Matlab, the attack was successfully implemented on two sets the traces. The result of attacks can be seen in Fig 18.

Comparing the result of two traces shows that the correlation coefficient of the protected design is decreased from 0.97 to 0.37. Although the area overhead the protection block is less than 0.001 of actual design, the level of resiliency, which it had provided is considerable that proves the effectiveness of the proposed idea. By changing the area of resiliency block, the level of protection can be modified too. This will result in power and area penalty, which is the case in all protection methods that has a trade-off between protection level and throughput, frequency, area and power. The final floorplan is shown in Fig. 19.

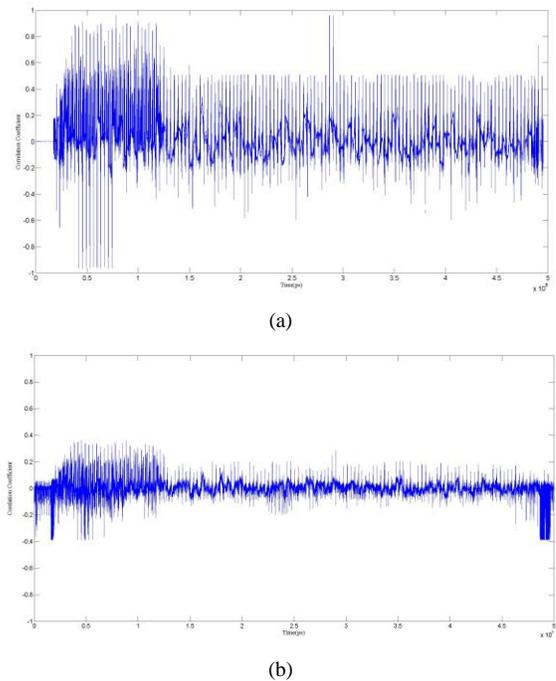

Fig. 18. Correlation Coefficient, (a) protected block, (b) non-protected block

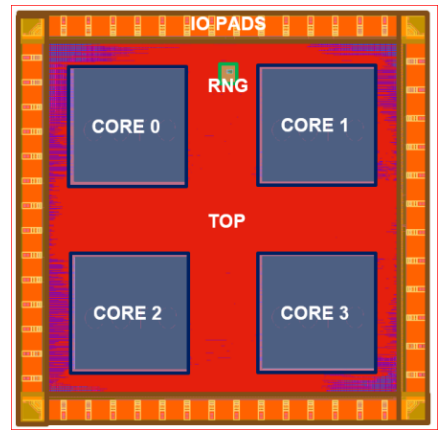

Fig. 19. Full-chip floorplan

TABLE II. COMPARSION WITH PREVIOUS RSA DESIGNS

| Year | Technology | Area | Power | Throughput | Frequency | Duration of RSA |
|---|---|---|---|---|---|---|
| 2011 [16] | 90nm | 861 gates | - | - | 432MHz | 118.7ms |
| 2011 [17] | 180nm | 107k gates | - | 578kbps | 450MHz | 1.79ms |
| 2011 [18] | - | - | - | - | - | - |
| 2013 [19] | 40nm | 804k tansistors | 350mW | - | 923MHz | 3.1mS |
| 2013 [20] | 65nm | 820kgates (1.925*3 um$^2$) | 400mW | - | 920MHz | 0.13ms |
| 2013 [21] | 180nm | 107k gates (2.73 mm$^2$) | 798mW | 121kbps | 125MHz | 8.44m |
| 2013 [22] | 180nm | 714676 um$^2$ | 40.3mW | 433.04kbps | - | - |
| OURS | 130nm | 20.25 mm$^2$ | 139 mW | 1250kbps | 222MHz | 0.8 ms |

## V. CONCLUSION

The 1024-bit RSA processor is fast and secure. Even though, the core of the RSA runs at 3.6ns, the processor clock periods is 4.5 ns due to the core library setup time. Consequently, the frequency is 222 MHz. The parallel core and high radix Montgomery multiplication reduces the total number of clock cycles required, thereby reducing the RSA computational time to 1.2 ms. Implemented in IBM 130 nm technology, the processor (excluding the IO pads) is 20.2 sq.mm in area and the power consumption is about 139 mW. The power resilience block adds very less area overhead of 0.02% and improves the resiliency by reducing the correlation coefficient from 0.9 to 0.4. In conclusion, the high radix k-partition implementation of Montgomery multiplication reduces the number of computation cycles considerably. Pipelining and interleaving the single core Montgomery multiplication would provide fast computation in lesser area. This approach could be explored further. The Random Number Generator used for power resiliency can be up-sized and is reusable across crypto-processors.

## VI. TEAM ORGANIZATION

At some point of time one or the other team member did each type of work. So everyone rotated the work among themselves and worked on all topics in all phases.

## REFERENCES

[1] Daly, Alan, and William Marnane. "Efficient architectures for implementing montgomery modular multiplication and RSA modular exponentiation on reconfigurable logic." Proceedings of the 2002 ACM/SIGDA tenth international symposium on Field-programmable gate arrays. ACM, 2002.

[2] Bayam, K.A.; Ors, B., "Differential Power Analysis resistant hardware implementation of the RSA cryptosystem," Circuits and Systems, 2008. ISCAS 2008. IEEE International Symposium on, vol., no., pp.3314,3317, 18-21 May 2008

[3] Cohen, A.E.; Parhi, K.K., "Architecture Optimizations for the RSA Public Key Cryptosystem: A Tutorial," Circuits and Systems Magazine, IEEE , vol.11, no.4, pp.24,34, Fourthquarter 2011

[4] Hongjun Wang; Zhiwen Song; Xiaoyu Niu; Qun Ding, "Key generation research of RSA public cryptosystem and Matlab implement," Sensor Network Security Technology and Privacy Communication System (SNS & PCS), 2013 International Conference on , vol., no., pp.125,129, 18-19 May 2013

[5] Koç, C.K.; Acar, Tolga; Kaliski, B.S., Jr., "Analyzing and comparing Montgomery multiplication algorithms," Micro, IEEE , vol.16, no.3, pp.26,33, Jun 1996

[6] Néto, J.C.; Ferreira Tenca, A.; Ruggiero, W.V., "A Parallel and Uniform k-Partition Method for Montgomery Multiplication," Computers, IEEE Transactions on , vol.63, no.9, pp.2122,2133, Sept. 2014

[7] Tokunaga, C.; Blaauw, D., "Secure AES engine with a local switched-capacitor current equalizer," Solid-State Circuits Conference - Digest of Technical Papers, 2009. ISSCC 2009. IEEE International , vol., no., pp.64,65,65a, 8-12 Feb. 2009

[8] O. Kömmerling and M. G. Kuhn, "Design principles for tamper resistant smartcard processors," in Proceedings of the USENIX Workshop on Smartcard Technology, Chicago, Illinois, USA, May 1999, pp. 9–20.

[9] K. Gandolfi, C. Mourtel, and F. Olivier, "Electromagnetic analysis: Concrete results," in Proceedings of the 3rd International Workshop on Cryptographic Hardware and Embedded Systems (CHES), ser. Lecture Notes in Computer Science, Ç. K. Koç, D. Naccache, and C. Paar, Eds., vol. 2162. Paris, France: Springer-Verlag, May 13-16 2001, pp. 255–265.

[10] J.-J. Quisquater and D. Samyde, "Electromagnetic analysis (EMA): Measures and counter-measures for smard cards," in Proceedings of the International Conference on Research in Smart Cards: Smart Card Programming and Security (E-smart), ser. Lecture Notes in Computer Science, I. Attali and T. Jensen, Eds., vol. 2140. Cannes, France: Springer-Verlag, September 19-21 2001, pp. 200–210.

[11] P. Kocher, J. Jaffe, and B. Jun, "Differential power analysis," in Advances in Cryptology: Proceedings of CRYPTO'99, ser. Lecture Notes in Computer Science, M. Wiener, Ed., vol. 1666. Santa Barbara, CA, USA: Springer-Verlag, August 15-19 1999, pp. 388–397.

[12] A. Shamir and E. Tromer, "Acoustic cryptanalysis," Preliminary proof-of-concept presentation, 2004, http://www.wisdom.weizmann.ac.il/~tromer/acoustic/.

[13] Kocher, Paul C. "Timing attacks on implementations of Diffie-Hellman, RSA, DSS, and other systems." In Advances in Cryptology—CRYPTO'96, pp. 104-113. Springer Berlin Heidelberg, 1996.

[14] Amaki, T.; Hashimoto, M.; Onoye, T., "An oscillator-based true random number generator with jitter amplifier," Circuits and Systems (ISCAS), 2011 IEEE International Symposium on , vol., no., pp.725,728, 15-18 May 2011

[15] Jen-Wei Lee; Szu-Chi Chung; Hsie-Chia Chang; Chen-Yi Lee, "Processor with side-channel attack resistance," Solid-State Circuits Conference Digest of Technical Papers (ISSCC), 2013 IEEE International , vol., no., pp.50,51, 17-21 Feb. 2013

[16] Miyamoto, A.; Homma, N.; Aoki, T.; Satoh, A., "Systematic Design of RSA Processors Based on High-Radix Montgomery Multipliers," Very Large Scale Integration (VLSI) Systems, IEEE Transactions on , vol.19, no.7, pp.1136,1146, July 2011

[17] Sutter, G.D.; Deschamps, J.; Imana, J.L., "Modular Multiplication and Exponentiation Architectures for Fast RSA Cryptosystem Based on

[18] Jiang Huiping; Yang Guosheng, "Resistant against power analysis for a fast parallel high-radix RSA algorithm,"Electric Information and Control Engineering (ICEICE), 2011 International Conference on , vol., no., pp.1668,1671, 15-17 April 2011

[19] Devlin, B.; Ikeda, M.; Ueki, H.; Fukushima, K., "Completely self-synchronous 1024-bit RSA crypt-engine in 40nm CMOS," Solid-State Circuits Conference (A-SSCC), 2013 IEEE Asian , vol., no., pp.309,312, 11-13 Nov. 2013

[20] Shuai Wang; Jun Han; Yang Li; Yifan Bo; Xiaoyang Zeng, "A 920MHz quad-core cryptography processor accelerating parallel task processing of public-key algorithms," Custom Integrated Circuits Conference (CICC), 2013 IEEE , vol., no., pp.1,4, 22-25 Sept. 2013

[21] da Costa, C.A.; Moreno, R.L.; Carpinteiro, O.S.A.; Pimenta, T.C., "A 1024 bit RSA coprocessor in CMOS," Microelectronics (ICM), 2013 25th International Conference on , vol., no., pp.1,4, 15-18 Dec. 2013

[22] Shiann-Rong Kuang; Jiun-Ping Wang; Kai-Cheng Chang; Huan-Wei Hsu, "Energy-Efficient High-Throughput Montgomery Modular Multipliers for RSA Cryptosystems," Very Large Scale Integration (VLSI) Systems, IEEE Transactions on , vol.21, no.11, pp.1999,2009, Nov. 2013

[23] A. Amarnath, J. Bagherzadeh, J. Tan and R. G. Dreslinski, "3DTUBE: A Design Framework for High-Variation Carbon Nanotube-based Transistor Technology," *2019 IEEE/ACM International Symposium on Low Power Electronics and Design (ISLPED)*, Lausanne, Switzerland, 2019, pp. 1-6, doi: 10.1109/ISLPED.2019.8824874.

[24] J. Bagherzadeh and V. Bertacco, "3DFAR: A three-dimensional fabric for reliable multi-core processors," *Design, Automation & Test in Europe Conference & Exhibition (DATE), 2017*, Lausanne, 2017, pp. 310-313, doi: 10.23919/DATE.2017.7927006.